\documentclass[prl,aps,superscriptaddress,twocolumn]{revtex4-1}
\setcitestyle{super}
\usepackage{amsfonts,amssymb,amscd,amsthm}
\usepackage{graphicx}
\usepackage{threeparttable}
\usepackage{multirow}
\usepackage{mathrsfs}
\usepackage[intlimits]{amsmath}
\usepackage{array}
\usepackage{xcolor}
\usepackage[colorlinks, citecolor=red]{hyperref}
\usepackage{lineno}
\usepackage{etoolbox}
\usepackage{color,soul}

\definecolor{blue}{rgb}{0,0,1}
\definecolor{red}{rgb}{1,0,0}
\definecolor{green}{rgb}{0,1,0}

\newcommand{\ket}[1]{\ensuremath{\left|#1\right\rangle}}

\begin{document}

\title{Tantalum airbridges for scalable superconducting quantum processors}

\author{Kunliang Bu}
\thanks{These three authors contributed equally to this work.}
\affiliation{Tencent Quantum Laboratory, Tencent, Shenzhen, Guangdong 518057, China}

\author{Sainan Huai}
\thanks{These three authors contributed equally to this work.}
\affiliation{Tencent Quantum Laboratory, Tencent, Shenzhen, Guangdong 518057, China}

\author{Zhenxing Zhang}
\thanks{These three authors contributed equally to this work.}
\affiliation{Tencent Quantum Laboratory, Tencent, Shenzhen, Guangdong 518057, China}

\author{Dengfeng Li}
\affiliation{Tencent Quantum Laboratory, Tencent, Shenzhen, Guangdong 518057, China}

\author{Yuan Li}
\affiliation{Tencent Quantum Laboratory, Tencent, Shenzhen, Guangdong 518057, China}

\author{Jingjing Hu}
\affiliation{Tencent Quantum Laboratory, Tencent, Shenzhen, Guangdong 518057, China}

\author{Xiaopei Yang}
\affiliation{Tencent Quantum Laboratory, Tencent, Shenzhen, Guangdong 518057, China}

\author{Maochun Dai}\email{maochundai@tencent.com}
\affiliation{Tencent Quantum Laboratory, Tencent, Shenzhen, Guangdong 518057, China}

\author{Tianqi Cai}\email{tianqicai@tencent.com}
\affiliation{Tencent Quantum Laboratory, Tencent, Shenzhen, Guangdong 518057, China}

\author{Yi-Cong Zheng}\email{yicongzheng@tencent.com}
\affiliation{Tencent Quantum Laboratory, Tencent, Shenzhen, Guangdong 518057, China}

\author{Shengyu Zhang}
\affiliation{Tencent Quantum Laboratory, Tencent, Shenzhen, Guangdong 518057, China}

\begin{abstract}

The unique property of tantalum (Ta), particularly its long coherent lifetime in superconducting qubits and its exceptional resistance to both acid and alkali, makes it promising for superconducting quantum processors. It is a notable advantage to achieve high-performance quantum processors with neat and unified fabrication of all circuit elements, including coplanar waveguides (CPW), qubits, and airbridges, on the tantalum film-based platform. Here, we propose a reliable tantalum airbridges with separate or fully-capped structure fabricated via a novel lift-off method, where a barrier layer with aluminium (Al) film is first introduced to separate two layers of photoresist and then etched away before the deposition of tantalum film, followed by cleaning with piranha solution to remove the residual photoresist on the chip. We characterize such tantalum airbridges as the control line jumpers, the ground plane crossovers and even coupling elements. They exhibit excellent connectivity, minimal capacitive loss, effectively suppress microwave and flux crosstalk and offer high freedom of coupling. Besides, by presenting a surface-13 tunable coupling superconducting quantum processor with median $T_1$ reaching above 100 $\mu$s, the overall adaptability of tantalum airbridges is verified. The median single-qubit gate fidelity shows a tiny decrease from about 99.95\% for the isolated Randomized Benchmarking to 99.94\% for the simultaneous one. This fabrication method, compatible with all known superconducting materials, requires mild conditions of film deposition compared with the commonly used etching and grayscale lithography. Meanwhile, the experimental achievement of non-local coupling with controlled-Z (CZ) gate fidelity exceeding 99.2\% may further facilitate quantum low-density parity check (qLDPC) codes, laying a foundation for scalable quantum computation and quantum error correction with entirely tantalum elements.
\end{abstract}

\maketitle

\section{Introduction}\label{Sec1}

A high-performance superconducting quantum processor plays key role in landing a universal quantum computation. Various factors have impact on the performance of a quantum chip, including the material platforms~\cite{place2021new}, the introduced two-level systems (TLSs)~\cite{muller2015interacting, de2020two, thorbeck2022tls} and the low temperature measurement environment~\cite{koch2007charge, krantz2019quantum}. An appropriate material is critical to achieve a neat fabrication process as well as a high-performance quantum chip. Recently, tantalum (Ta) has obtained significant attentions as a new material in superconducting qubit systems owing not only to its long coherent lifetime exceeding 500 $\mu$s~\cite{wang2022towards, place2021new, ganjam2023surpassing}, but also to its superior stability compared to the commonly used superconducting materials like aluminium (Al) or niobium (Nb)~\cite{chen2014fabrication, murthy2022developing, altoe2022localization, verjauw2021investigation}.

Airbridges, a kind of free-standing metallic crossovers with separate or fully-capped structure, are widely used in superconducting quantum processors, functioning as a medium to enhance circuit freedom and density via its spatial architecture capability, as well as suppressing slotline modes and improving coherence for both coplanar waveguide (CPW) resonators and qubits via ground plane connection~\cite{chen2014fabrication, dunsworth2018method, abuwasib2013fabrication, jin2021microscopic, girgis2006fabrication, janzen2022aluminum, sun2022fabrication, stavenga2023lower}. Delicate fabrication processes have been developed to construct the airbridges on chip utilizing the etching~\cite{chen2014fabrication}, the vapor HF release~\cite{dunsworth2018method}, multi-layer photoresist lift-off~\cite{girgis2006fabrication}, and grayscale lithography~\cite{sun2022fabrication, stavenga2023lower}. However, most of existing techniques are applicable only to the aluminium, and the progresses to construct tantalum airbridges are lacking.

In this work, we propose and demonstrate a novel lift-off method to fabricate airbridges with two layers of photoresist, which is compatible with all known superconducting materials, particularly suitable for tantalum. A barrier layer of aluminum film is initially introduced as a sacrificing layer and partially removed prior to the film deposition process. The second layer of photoresist is subsequently applied while preserving the integrity of the airbridge scaffolds. Notably, tantalum film-based quantum chips with tantalum airbridges can be further cleaned via the piranha solution to eliminate the residual photoresist and particles. We evaluate the performance and versatility of these tantalum airbridges as the control line jumpers and the ground plane crossovers through connectivity test, measurement of internal quality factor $Q_i$ for CPW resonators and characterization of microwave/flux crosstalk. The results show the performance of tantalum airbridges with minimal capacitive loss per bridge [$\left(3.84 \pm 0.08\right)\times 10^{-9}$] and negligible crosstalk interference (median $1.4 \times 10^{-4}$ for flux crosstalk and -45 dB for microwave crosstalk). To further illustrate the scalability and adaptability of tantalum airbridges in multiqubit fabrication processes, we present a tantalum film-based surface-13 tunable coupling superconducting quantum processor equipped with fully-capped tantalum airbridges with a median $T_1$ exceeding 100 $\mu$s. We perform the Randomized Benchmarking (RB) for all thirteen qubits, and the median single qubit gate fidelity shows a tiny decrease from about 99.95\% for isolated-RB to 99.94\% for simultaneous-RB. Furthermore, considering the advantage of non-local coupling for fault-tolerant quantum computation with quantum low-density parity check (qLDPC) codes~\cite{breuckmann2021quantum, fujiwara2010entanglement, strikis2023quantum, kovalev2013fault, kovalev2012fault, panteleev2021degenerate, panteleev2021quantum} and quantum simulation~\cite{yanay2022mediated, liu2007entanglement, song201710, zhao2020switchable}, we investigate direct qubit-qubit connections via tantalum airbridges by offering extra spacial coupling freedom. Our experiments demonstrate that tantalum airbridges effectively facilitate coupling between qubits without compromising coherence time, with two-qubit controlled-Z (CZ) gate fidelity achieving 99.2\% via RB~\cite{knill2008randomized}.

\section{Fabrication procedure}\label{Sec2}

\begin{figure}[t]
\includegraphics{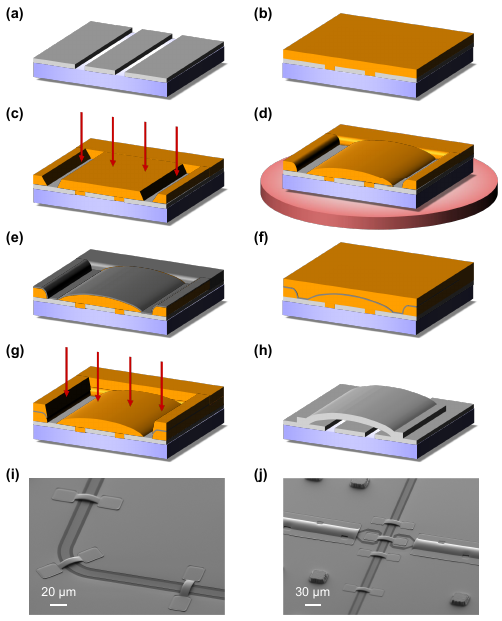}
\caption{(a)-(h) Schematic diagram of key processes in the fabrication of superconducting tantalum airbridges, with substrate shown in violet, tantalum in light gray, aluminium in dark gray and photoresist in orange. (a) Fabrication of the CPW layer with a 200 nm thick $\alpha$-phase tantalum film on the sapphire substrate. (b)-(d) Formation of the photoresist scaffold and base layer for airbridges with (b) patterning, (c) exposure and (d) reflow of the first layer of photoresist. The red arrows in (c) denote the ultraviolet light, while the red disk in (d) represents baking at $140^{\circ}$C. (e) Deposition of a barrier layer with 25 nm thick aluminium. (f)(g) Formation of the bridge layer with (f) spin coating and (g) exposure of the second layer of photoresist. (h) Deposition of the tantalum airbridge with the niobium functioned as the seed layer after the ion milling of the CPW to ensure an electrical connection. (i)(j) SEM images of tantalum airbridges with separate and fully-capped structures connecting the ground planes or two CPW center lines by lift-off method.}
\label{fig:Fig1}
\end{figure}

The detailed fabrication procedure of the tantalum airbridges is illustrated schematically in Fig.~\ref{fig:Fig1}(a)-(h). The CPW was first fabricated on a sapphire substrate with a 200 nm thick $\alpha$-phase tantalum film~\cite{place2021new, wang2022towards}. Due to exposure to the atmosphere, a native oxide layer was naturally grown on the top surface. The base layer for airbridges was then developed after subjecting the first layer of photoresist (SPR220) to ultraviolet exposure, followed by $140^{\circ}$C reflow process to form an arch for mechanical stability, which ensures a continuous connection between suspending crossovers and base layers~\cite{chen2014fabrication}. Next, a 25 nm thick aluminum film was deposited as a barrier layer between two layers of photoresist. It should be careful to select an appropriate thickness for this barrier layer. During high-speed spin coating, a too-thin aluminum film could lead to cracks, while a too-thick film could induce instability in small structures, especially for fully-capped airbridges during subsequent development processes. To define the bridge layer, a second layer of photoresist (either positive or negative type; SPR220 in our case) was covered on the barrier layer to construct the main structure of the airbridges. After that, the barrier layer (sacrificing layer) was etched away using TMAH (Tetramethylammonium hydroxide) alkali developer solution. Following this step, we deposited a 400 nm tantalum film onto the obtained structure in an magnetron sputtering to form the bridge layer. Prior to the deposition, we performed an $in~situ$ argon ion milling for 2 minutes under the pressure of $1.0 \times 10^{-4}$ Torr to remove any native oxide present on base tantalum surface and ensure proper electrical contact. After tantalum deposition, we applied ultrasonic treatment with deionized water to the samples, which effectively destroyed the fragile connection between airbridge base layers and nearby tantalum film on the sidewall of the photoresist. Specially, another dip in a Transene Aluminum Etchant Type A solution could be selected to remove the aluminum residual for tantalum airbridges. Finally, the sample was dipped in the PG Remover to go through lift-off process, followed by the plasma cleaning. The corresponding scanning electron microscope (SEM) images of the tantalum airbridges can be found in Fig.~\ref{fig:Fig1}(i)-(j).

The presence of an aluminum barrier layer is crucial in the proposed lift-off method for several reasons. Firstly, it effectively prevents the mixing of two layers of photoresist used in the recipe, as most types of photoresist are susceptible to blending, leading to pattern degradation during the spin coating of the second layer. Secondly, it provides well protection for the arc-shaped scaffold created by the first layer of photoresist against exposure to the second one and eliminates the potential damage caused by subsequent alkali development. Thirdly, after undergoing argon ion milling, a large amount of stubborn residual photoresist on sidewalls remains unremovable by solvents~\cite{chen2014fabrication}. Fortunately, the barrier layer is usually over etched, leaving a small gap between two layers of photoresist that effectively disconnects the stubborn residue and benefits a clean surface.

In practice, the etching~\cite{chen2014fabrication} and grayscale lithography methods~\cite{sun2022fabrication, stavenga2023lower} can also be employed for the fabrication of tantalum airbridges in principle. However, they may impose more stringent conditions on the tantalum film deposition compared to the proposed lift-off method. Considering the poor temperature resistance of photoresist, tantalum can only be deposited at room or low temperature assisted by utilizing niobium as a seed layer for airbridges~\cite{jones2023grain, dorranian2011effects}. Meanwhile, corrugations may occur on the photoresist due to thermal effects during the sputtering of tantalum. Conversely, such thermal effects are absent in airbridge areas for the lift-off method since they contact with the lower substrates, thus greatly simplify the process.

\section{Experimental results}\label{Sec3}

\subsection{Characteristics}

\begin{figure}
\includegraphics{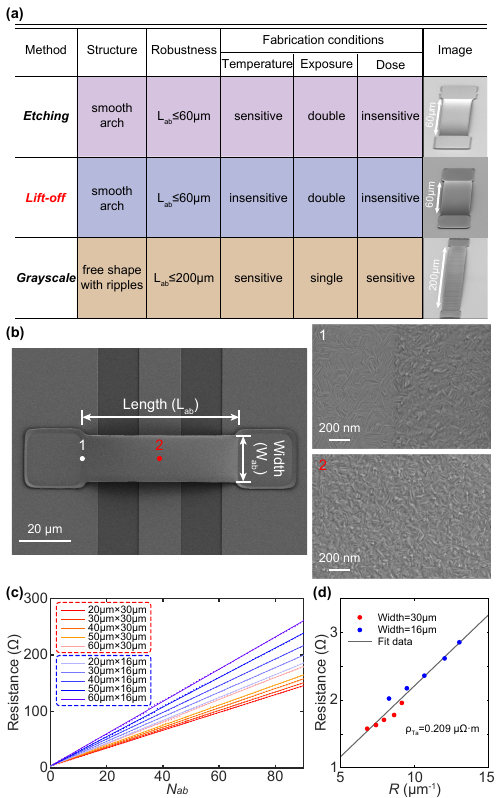}
\caption{(a) Comparison among three fabrication methods for airbridges, where the ``robustness" in the table denotes the length limitation to ensure intact airbridges with a fixed height of 3 $\mu$m and arbitrary width, while the ``temperature" in ``fabrication conditions" represents a critical requirement for temperature control during film deposition. (b) SEM images of a single tantalum airbridge with the full structure shown in left panel and zoom-in structures shown in right panel. The length ($L_{ab}$) and width ($W_{ab}$) of the airbridge are defined as the white arrow in the left panel. (c) Measured resistance versus number of airbridges ($N_{ab}$) with legend ``$L_{ab}$ $\times$ $W_{ab}$'' of airbridges defined in (b). (d) Resistance of single unit versus resistance ratio $R$. The red and blue dots represent raw data acquired from airbridges with width of 30 $\mu$m and 16 $\mu$m respectively.}
\label{fig:Fig2}
\end{figure}

To obtain the appropriate design patterns and limits for fabricating airbridges, we systematically explored various lengths and widths using the etching, grayscale lithography, and the proposed lift-off. For airbridges with a height of 3 $\mu$m, the maximum length for achieving structural stability is limited to be 60 $\mu$m based on the etching and lift-off; however, when employing grayscale lithography, this limit can be extended to 200 $\mu$m, as illustrated in Fig.~\ref{fig:Fig2}(a)(b). This discrepancy arises from the observation that beyond the length of 60 $\mu$m, plateaus appear in the profiles of photoresist scaffolds for those using etching or lift-off but not for grayscale lithography. We extracted the profile of photoresist sidewalls from a reflow surface and simulated scaffold profiles, which showed consistent evolution with experimental data (see Supplementary Materials). This approach allows us to estimate the maximum length of an airbridge without conducting further experiments. In Fig.~\ref{fig:Fig2}(a), we compare these three methods, finding that the lift-off method is robust and fabrication friendly, especially in constructing tantalum airbridges.

In addition, we carefully analyzed the crystal morphology of tantalum airbridges, as depicted in the right panel of Fig.~\ref{fig:Fig2}(b). The tantalum film on the base layer of airbridges shows a crystal grain similar to $\alpha$ phase, while the tantalum film on top of airbridges differs from typical $\alpha$ phase. The distinct grain structures observed at different positions can be attributed to variations in growth bases for the tantalum film; specifically, $\alpha$-phase tantalum (amorphous photoresist) serves as the base for tantalum film grown on the base (bridge) layer. Consequently, our results suggest that both deposition environment and base materials significantly influence the phase of tantalum films. 

Figure~\ref{fig:Fig2}(c) demonstrates the measurement of resistance for tantalum airbridges chains to analyze the connectivity with varied lengths and widths. It is observed that the total resistance of tantalum airbridges scales linearly with number of airbridges. Here, the slope of fitted line represents the resistance of a single unit, comprising a tantalum airbridge and a conducting pad, and we define a total resistance ratio $R$ for each single unit as 
\begin{equation}\label{eq: eq1}
\begin{split}
R = \frac{L_{ab}}{(W_{ab} \times t_{ab})} + \frac{L_{pad}}{(W_{pad} \times t_{pad})},
\end{split}
\end{equation}
where $L_{ab}$ ($L_{pad}$), $W_{ab}$ ($W_{pad}$) and $t_{ab}$ ($t_{pad}$) denote the length, width and thickness of the airbridge (conducting pad) respectively. The slope of the fitted line is then extracted and plotted against the total ratio, as depicted in Fig.~\ref{fig:Fig2}(d). This analysis yields a resistivity value of approximately 0.21 $\mu \Omega \cdot$m for each single unit. The small resistivity indicates a well contact using tantalum airbridges.

\subsection{$Q_i$ measurement}

\begin{figure}
\includegraphics{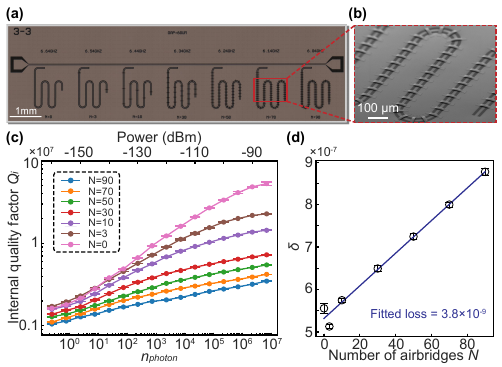}
\caption{(a) Optical micrograph of one resonator chip with varied tantalum airbridges. The number of airbridges increases from $N=0$ to $N=90$ and the resonance frequencies are uniformly distributed between 6.04 GHz and 6.64 GHz. (b) A zoom-in SEM image of the arranged tantalum airbridges circled by the red rectangle in (a). (c) Internal quality factor $Q_i$ versus average photon population $\left < n_p \right >$ for each resonator. The error bars represent the uncertainty of $Q_i$ in the fitting process of $S$ parameters. Notice that the overall $Q_i$ values tend to decline as the number of airbridges increases. (d) Loss tangent ($1/Q_i$) versus number of airbridges. The fitted slope of line represents the added capacitive loss caused by a single airbridge to the CPW resonator.}
\label{fig:Fig3}
\end{figure}

\begin{figure*}
\includegraphics{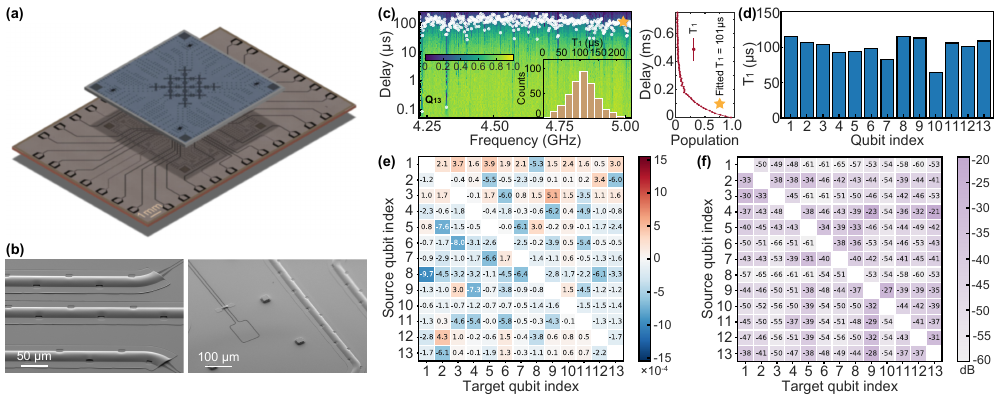}
\caption{(a) Optical micrograph of the surface-13 flip-chip processor. Here, the top chip is primarily used for qubit layer containing both Josephson junctions and shunt capacitors, while the bottom chip comprises readout resonators and control lines. Tantalum airbridges with fully-capped structure are incorporated over each control line to minimize crosstalk. (b) SEM images of the tantalum airbridges with fully-capped structure in (a). (c) Characterization of $T_1$ measurement varied with qubit frequency for $Q_{13}$. The inset illustrates the histogram with a median $T_1$ of 109 $\mu$s. A representative single $T_1$ measurement result is shown in the right panel. (d) Median $T_1$ results for all thirteen qubits. Most qubits exhibit a median $T_1$ exceeding 100 $\mu$s. (e)(f) Measurement of (e) flux crosstalk and (f) microwave crosstalk with tantalum airbridges utilizing the fully-capped structure.}
\label{fig:Fig4}
\end{figure*}

To quantify the added capacitive loss from the tantalum airbridges at low temperature ($\sim$20 mK in the experiment), we fabricated resonator chip containing seven $\lambda/4$ CPW resonators with various number of airbridges ranging from $N=0$ to $N=90$, as illustrated in Fig.~\ref{fig:Fig3}(a)~\cite{chen2014fabrication, janzen2022aluminum}. The airbridges were designed with a length of 60 $\mu$m and width of 16 $\mu$m, automatically arranged at equal distances using our home made electronic design automation (EDA) tool for each resonator. A zoom-in image displaying the arrangement of the airbridges is shown in Fig.~\ref{fig:Fig3}(b). Careful design considerations were taken into account for the resonators to ensure an appropriate line width (also coupling quality factor $Q_c$) that facilitates subsequent measurement of internal quality factor ($Q_i$). Furthermore, the resonance frequencies were uniformly distributed between 6.04 GHz and 6.64 GHz to avoid potential microwave electrical crosstalk~\cite{janzen2022aluminum}.

The internal quality factor $Q_i$ was determined by measuring and fitting the $S$ parameters of these resonators with varied drive power, allowing for a range of photon populations $\left < n_p \right >$ in the resonator from single photon to high power photons\cite{megrant2012planar}. Figure~\ref{fig:Fig3}(c) presents representative $Q_i$ data for resonators with different numbers of airbridges. It is evident that at both low and high power, the trend of $Q_i$ exhibits two plateaus corresponding to TLS-determined behavior at around single photon level and TLS-saturated behavior at high power level. For tantalum resonator without added airbridges, the internal quality factor is measured to be $2.0 \times 10^{6}$ at single photon power and increases to $5.3 \times 10^{7}$ at high photon power, consistent with the theoretical variation tendency. Similar trends are observed for tantalum resonators with a few tantalum airbridges. However, the overall $Q_i$ values appear to decrease with an increasing number of airbridges, indicating additional loss introduced by tantalum airbridges that can be further characterized by fitting the loss tangent ($1/Q_i$) data. By observing the linear relationship between number of airbridges and the loss tangent, we estimate an added loss per bridge of approximately $\left(3.84 \pm 0.08\right)\times 10^{-9}$, demonstrating comparable low-loss and high-quality performance of tantalum airbridges to aluminum airbridges~\cite{chen2014fabrication, janzen2022aluminum}.

\section{Discussions}\label{Sec4}
 
The tantalum airbridges with lift-off method have been validated with a robust fabrication process, well connectivity, and minimal additional loss to the resonators. Consequently, they hold immense potential for widespread utilization in quantum processors~\cite{arute2019quantum, wu2021strong, shi2022observing}. To further demonstrate their capabilities, we explore two arenas: crosstalk suppression using fully-capped tantalum airbridges and direct coupling element employing well-designed tantalum airbridges.

\subsection{Crosstalk suppression}

We investigated the usage of tantalum airbridges in a tunable coupling quantum processor based on a tantalum film and flip-chip architecture, which comprises 13 qubits and 16 couplers, as the schematic diagram shown in Fig.~\ref{fig:Fig4}(a)~\cite{conner2021superconducting, rosenberg20173d}. All computational qubits are frequency-tunable transmon qubits, with grounded couplers enabling continuous adjustment of the effective coupling strength between qubits~\cite{yan2018tunable, li2020tunable}. The design and optimization of the tantalum airbridges in this chip were achieved through a parametric EDA method, where the center line of the CPW can be automatically identified for the chip layout with wiring, followed by adding the airbridges accordingly. The fabrication followed the proposed lift-off method depicted in Fig.~\ref{fig:Fig1} (the complete fabrication process of this 13-qubit quantum chip can be found in Supplementary Materials). Due to the resistance to acid and alkaline solutions, we could immerse the chip into piranha solution to remove residual photoresist and eliminate the oxygen layer on the substrate after fabricating the tantalum airbridges – an option typically unavailable for aluminum airbridges. SEM images showcasing fully-capped tantalum airbridges wrapped around control lines are presented in Fig.~\ref{fig:Fig4}(b).

\begin{figure*}
\includegraphics{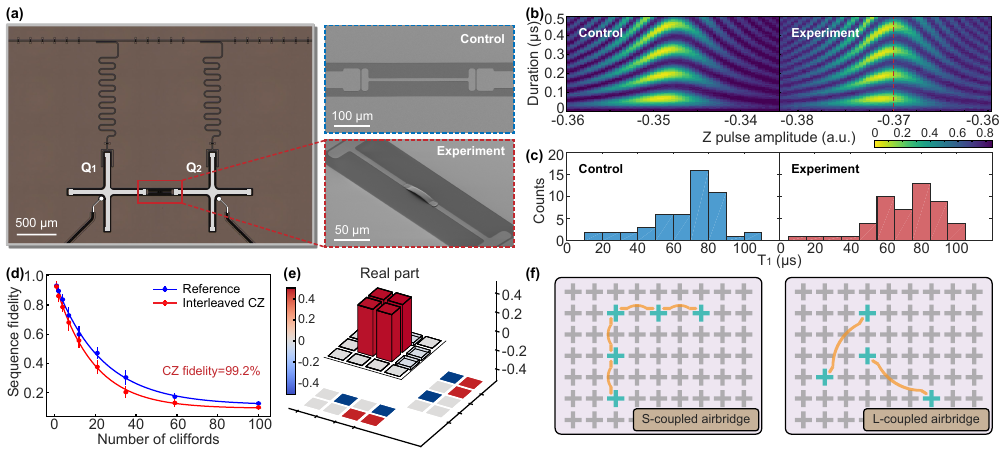}
\caption{(a) Optical micrograph of the two-qubit superconducting quantum chip coupled via the tantalum airbridge. Zoom-in images of the coupling for both control group and experimental group are shown in the right panel respectively. (b) Characterization of effective coupling strength. The qubits in both groups are initially prepared in the $\ket{01}$ state (or $\ket{10}$ state), followed by fine-tuning the qubit frequencies to achieve resonance. The red dotted line in the experimental group represents the maximum resonance position where we extract the coupling strength to be 3.9 MHz. (c) $T_1$ distribution for both control group and experimental group. The coupling capacitor via airbridge seems to have little effect on qubit decoherence. (d) Interleaved-RB of diabatic CZ gate for the experimental group, achieving consistent gate fidelity of 99.2\% compared to the control group. (e) Preparation of Bell state $\ket{\psi}=\left(\ket{01}+\ket{10} \right)/\sqrt{2}$ using CZ gate in (d) with state fidelity measured to be 99.7\% via QST. (f) Schematic diagrams of the non-local coupling mechanism via tantalum airbridges in superconducting quantum processors. Here we consider two potential airbridge structures for coupling. Left panel: S-coupled airbridge realizing neighboring or next-neighboring qubit-qubit connection. Right panel: L-coupled airbridge realizing long distance qubit-qubit connection.}
\label{fig:Fig5}
\end{figure*}

The fundamental performance of this flip-chip processor was characterized, with the results illustrated in Fig.~\ref{fig:Fig4}(d). It can be observed that most qubits exhibit a median $T_1$ exceeding 100 $\mu$s extracted from the $T_1$ distribution varied with qubit frequencies, see Fig.~\ref{fig:Fig4}(c) as an example (Supplementary Materials provide complete data for other qubits). Meanwhile, a representative single $T_1$ measurement is depicted in the right panel of Fig.~\ref{fig:Fig4}(c). Notably, when TLS occurs at certain frequencies, a significant decline in population may occur, bringing detrimental effect to qubits~\cite{ku2005decoherence}. Moreover, we evaluated the microwave and flux crosstalk between each control line on the chip. Previous studies have effectively mitigated such impact by introducing a crosstalk matrix method~\cite{reed2013entanglement, barrett2023learning, dai2021calibration, li2020tunable, nuerbolati2022canceling, yan2023calibration}. Nevertheless, this approach takes tremendous time with growing number of qubits due to the quadratic scaling of experiments, thus compensating for crosstalk between control lines becomes challenging~\cite{barrett2023learning}. To eliminate crosstalk without relying on such compensation, we realized tantalum airbridges with a fully-capped structure on all control lines. Fully-capped airbridges enable complete coverage and shield microwave electrical signals to significantly suppress crosstalk~\cite{arute2019quantum}. The corresponding measurement results are presented in Fig.~\ref{fig:Fig4}(e) and (f), revealing negligible median flux crosstalk at 1.4$\times 10^{-4}$ and microwave crosstalk at -45 dB - one to two orders of magnitude lower than that observed using traditional separate airbridges~\cite{barrett2023learning, yan2023calibration}. To further assess the isolation of the fully-capped airbridges, we also benchmark the single qubit gate fidelity using both isolated and simultaneous RB. The median fidelity is found to be 99.95\% for isolated-RB and 99.94\% for simultaneous-RB. The tiny decrease ($1.3\times 10^{-4}$) indicates that our device exhibits low microwave crosstalk provided by the airbridges (see detailed data in the Supplementary Materials).

\subsection{Non-local coupling elements}

Furthermore, we discuss the coupling through airbridges, a rarely explored scenario before. Superconducting qubits can typically be coupled through direct coupling~\cite{wang2019experimental}, coupler coupling~\cite{arute2019quantum, li2020tunable}, resonator coupling~\cite{song201710}, or waveguide coupling~\cite{kannan2020waveguide, kockum2018decoherence, yin2023generation}. For instance, the well-known surface code requires nearest neighbor qubit coupling, rendering direct or coupler coupling suitable choices~\cite{fowler2012surface, google2023suppressing}. However, long-distance qubit-qubit connections or fully-connected architecture necessitates the utilization of resonator or waveguide coupling. The resonators provide various geometrical options for connecting qubits but they also impose limitations on chip size and scalability. Recently, qLDPC codes offer considerable promise in diminishing the overall resources. However, the low-density parity check matrix puts extra requirement on non-local (non-neighboring) qubit couplings for syndrome measurements. Here, we propose a method for connecting qubits using airbridges that effectively leverages spatial chip structures to achieve such non-local couplings.

Figure~\ref{fig:Fig5}(a) illustrates a two-qubit superconducting quantum chip with tantalum airbridge-coupling fabricated via the lift-off method. To benchmark the performance of the qubits coupled with the tantalum airbridge (experimental group), we designed another pair of qubits coupled with a complete grounded capacitor (control group), as shown in Fig.~\ref{fig:Fig5}(a). We first characterized the effective coupling strength between these two qubits using airbridge coupling as depicted in Fig.~\ref{fig:Fig5}(b). By tuning the frequency of one qubit to resonate with the other, the coupling strength can be extracted to be around 3.9 MHz, similar to that achieved in the control group. Next, we measured the coherence time of the qubits as displayed in Fig.~\ref{fig:Fig5}(c), finding that they are unlikely to be affected by introducing a tantalum airbridge for coupling, confirming its feasibility. Furthermore, the quality of such non-local coupling via an airbridge was conducted through a two-qubit CZ gate process. The flux pulses applied to each qubit were carefully designed with slowly changing waveforms and adjusted to generate resonance process for states $\ket{11}$ and $\ket{02}$ (or $\ket{20}$), followed by the accumulation of $\pi$ phase, thereby realizing standard diabatic CZ gate operation~\cite{xu2021realization, marxer2023long}. The fidelity for this two-qubit CZ gate was characterized through interleaved-RB, which yields 99.2\% as depicted in Fig.~\ref{fig:Fig5}(d), comparable to the one acquired from the control group. Meanwhile, we verified the entanglement capability by preparing Bell state with such CZ gate, with a state fidelity of 99.7\% via quantum state tomography (QST), as illustrated in Fig.~\ref{fig:Fig5}(e). Finally, we provide schematic diagrams in Fig.~\ref{fig:Fig5}(f) to illustrate the application of the proposed airbridge-coupling mechanism in the multiqubit quantum chip with non-local couplings. Through further improvement like the design of new bridge structures assisted with optimized lift-off or grayscale lithography, we believe that length of tantalum airbridges (especially L-coupled airbridges) can be further expanded for long-distance connectivity of quantum chip, advancing fault-tolerant quantum computation on qLDPC codes and quantum simulations.

In summary, we propose and develop a novel lift-off method for fabricating tantalum airbridges with a separate and fully-capped structure. We compare three fabrication methods including etching, grayscale lithography, and the proposed lift-off. We verify the robustness of our approach under varied fabrication conditions, with an emphasis on temperature control during film deposition. This aims to alleviate stringent device requirements. Furthermore, excellent performance and versatility of these tantalum airbridges as control line jumpers and ground plane crossovers are illustrated through connectivity tests, $Q_i$ measurement of CPW resonators, and crosstalk calibration. To demonstrate the scalability and adaptability of tantalum airbridges in multiqubit fabrication processes, we present a surface-13 tunable coupling superconducting quantum processor equipped with fully-capped tantalum airbridges based on tantalum films, with median $T_1$ exceeding 100 $\mu$s for most qubits, and consistent single-qubit gate fidelity for isolated-RB and simultaneous-RB. Finally, we explore the application of tantalum airbridges in non-local coupling, achieving measured two-qubit CZ gate fidelity of 99.2\% without sacrificing coherence time. Our results demonstrate the reliability of the lift-off method for fabricating tantalum airbridges and its potential broad applications in quantum computation and quantum error correction.


%

\bigskip

\end{document}


\title{Supplementary Materials: Tantalum airbridges for scalable superconducting quantum processors}

\author{Kunliang Bu}
\thanks{These three authors contributed equally to this work.}
\affiliation{Tencent Quantum Laboratory, Tencent, Shenzhen, Guangdong 518057, China}

\author{Sainan Huai}
\thanks{These three authors contributed equally to this work.}
\affiliation{Tencent Quantum Laboratory, Tencent, Shenzhen, Guangdong 518057, China}

\author{Zhenxing Zhang}
\thanks{These three authors contributed equally to this work.}
\affiliation{Tencent Quantum Laboratory, Tencent, Shenzhen, Guangdong 518057, China}

\author{Dengfeng Li}
\affiliation{Tencent Quantum Laboratory, Tencent, Shenzhen, Guangdong 518057, China}

\author{Yuan Li}
\affiliation{Tencent Quantum Laboratory, Tencent, Shenzhen, Guangdong 518057, China}

\author{Jingjing Hu}
\affiliation{Tencent Quantum Laboratory, Tencent, Shenzhen, Guangdong 518057, China}

\author{Xiaopei Yang}
\affiliation{Tencent Quantum Laboratory, Tencent, Shenzhen, Guangdong 518057, China}

\author{Maochun Dai}\email{maochundai@tencent.com}
\affiliation{Tencent Quantum Laboratory, Tencent, Shenzhen, Guangdong 518057, China}

\author{Tianqi Cai}\email{tianqicai@tencent.com}
\affiliation{Tencent Quantum Laboratory, Tencent, Shenzhen, Guangdong 518057, China}

\author{Yi-Cong Zheng}\email{yicongzheng@tencent.com}
\affiliation{Tencent Quantum Laboratory, Tencent, Shenzhen, Guangdong 518057, China}

\author{Shengyu Zhang}
\affiliation{Tencent Quantum Laboratory, Tencent, Shenzhen, Guangdong 518057, China}

\maketitle

\tableofcontents

\newpage

\section{Theoretical analyses and circuit design}\label{SecS1}

\subsection{Circuit Hamiltonian and quantization}

\begin{figure}
\includegraphics{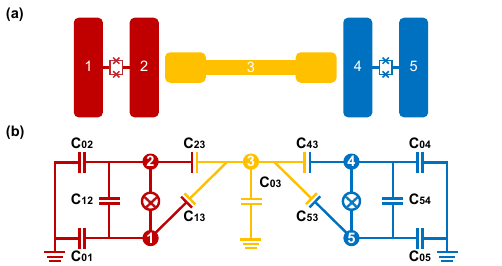}
\caption{(a) Schematic diagram of the two floating transmon qubits coupled via the airbridge. (b) The equivalent lumped-circuit representation of the airbridge-coupled architecture with ignoring $C_{14}, \, C_{15}, \, C_{24}, \, C_{25}$ in (a).}
\label{fig:FigS1}
\end{figure}

We conduct a theoretical investigation on the circuit Hamiltonian and quantization process of the airbridge-coupled superconducting quantum chip. The airbridge plays a crucial role in connecting the coupling capacitors, making this architecture analogous to the commonly-used qubit-qubit direct coupling scheme via a grounded capacitance~\cite{wang2019experimental}. To clearly present the design parameters, we provide an illustrative example below where two floating transmon qubits are coupled by means of an airbridge, as depicted in Fig.~\ref{fig:FigS1}. The Lagrangian of this architecture in terms of the node fluxes and the circuit elements can be directly written as~\cite{sete2021parametric, sete2021floating, marxer2023long, liang2023tunable}
\begin{equation}\label{eq: Lagrangian}
\begin{split}
L = T-U,
\end{split}
\end{equation}
where
\begin{equation}\label{eq: Lagrangian2}
\begin{split}
T &= \sum_{i=1}^{5} \frac{1}{2} C_{0i} \dot{\Phi}_{i}^2 + \sum_{i,j\neq 0} \frac{1}{2} C_{ij} \left(\dot{\Phi}_{i} - \dot{\Phi}_{j} \right)^2 \\
& \approx \frac{1}{2} C_{01} \dot{\Phi}_{1}^2 + \frac{1}{2} C_{02} \dot{\Phi}_{2}^2 + \frac{1}{2} C_{03} \dot{\Phi}_{3}^2 + \frac{1}{2} C_{04} \dot{\Phi}_{4}^2 + \frac{1}{2} C_{05} \dot{\Phi}_{5}^2 \\
&+ \frac{1}{2} C_{12} \left(\dot{\Phi}_{1} - \dot{\Phi}_{2} \right)^2 + \frac{1}{2} C_{23} \left(\dot{\Phi}_{2} - \dot{\Phi}_{3} \right)^2 \\
&+ \frac{1}{2} C_{54} \left(\dot{\Phi}_{5} - \dot{\Phi}_{4} \right)^2 + \frac{1}{2} C_{43} \left(\dot{\Phi}_{4} - \dot{\Phi}_{3} \right)^2 \\ 
&+ \frac{1}{3} C_{13} \left(\dot{\Phi}_{1} - \dot{\Phi}_{3} \right)^2 + \frac{1}{2} C_{53} \left(\dot{\Phi}_{5} - \dot{\Phi}_{3} \right)^2 \\
U &= -E_{J1}\cos(\phi_2 - \phi_1 + \phi_{01}) - E_{J2}\cos(\phi_4 - \phi_5 + \phi_{02}).
\end{split}
\end{equation}
Here, $C_{ij}$ are capacitances and $\Phi_k$ ($k=1 \sim 5$) are node fluxes; $\phi_k=2\pi \Phi_k/\Phi_0$ with $\Phi_0=h/2e$ the flux-quantum; $E_{Jm} = \sqrt{E_{Jsm}^2 + E_{Jlm}^2 + 2E_{Jsm}E_{Jlm}\cos(\phi_{em})}$ ($m=1, \, 2$) with $E_{Jsm}$ and $E_{Jlm}$ being the smaller and larger Josephson energies in SQUID $m$; $\phi_{0m} = \tan^{-1} \left[ \frac{E_{Jsm}-E_{Jlm}}{E_{Jsm}+E_{Jlm}} \tan(\frac{\phi_{em}}{2}) \right]$ ($m=1, \, 2$) with $\phi_{em}$ being the external flux bias through the SQUID. Notice that we have assumed that $C_{14}=C_{15}=C_{24}=C_{25}=0$ for simplicity in consideration of the real situation in this architecture. In practice, the canonical momentum $Q_i$ and the coordinate $\Phi_i$ conform to commutation relations, thereby the momentum can be given as $Q_i = \partial L/\partial \dot{\Phi}_i$ according to the Lagrangian in Eq.~(\ref{eq: Lagrangian}), and can be further expressed in the matrix form $\mathbf{Q} = \mathbf{C} \mathbf{\dot{\Phi}}$ as
\begin{equation}\label{eq: CapacitanceMatrix}
\begin{split}
\left(\begin{array}{ccccc} Q_1 \\ Q_2 \\ Q_3 \\ Q_4 \\ Q_5 \end{array} \right) = 
\left(\begin{array}{ccccc} 
C_{1S} & -C_{12} & -C_{13} & 0 & 0 \\
-C_{12} & C_{2S} & -C_{23} & 0 & 0 \\
-C_{13} & -C_{23} & C_{3S} & -C_{43} & -C_{53} \\
0 & 0 & -C_{43} & C_{4S} & -C_{54} \\
0 & 0 & -C_{53} & -C_{54} & C_{5S}
\end{array} \right)
\left(\begin{array}{ccccc} \dot{\Phi}_1 \\ \dot{\Phi}_2 \\ \dot{\Phi}_3 \\ \dot{\Phi}_4 \\ \dot{\Phi}_5 \end{array} \right),
\end{split}
\end{equation}
where $C_{1S}=C_{01}+C_{12}+C_{13}$, $C_{2S}=C_{02}+C_{12}+C_{23}$, $C_{3S}=C_{03}+C_{13}+C_{23}+C_{43}+C_{53}$, $C_{4S}=C_{04}+C_{43}+C_{54}$, and $C_{5S}=C_{05}+C_{53}+C_{54}$. Since the transmon qubits in Fig.~\ref{fig:FigS1} are the floating qubits, we introduce new variables as $\Phi_{1p/m} = \Phi_2 \pm \Phi_1$, $\Phi_{2p/m} = \Phi_4 \pm \Phi_5$ with the transformation matrix form $\mathbf{\tilde{\Phi}}=\mathbf{S}\mathbf{\Phi}$ as~\cite{sete2021parametric, sete2021floating}
\begin{equation}\label{eq: SMatrix}
\begin{split}
\left(\begin{array}{ccccc} \Phi_{1p} \\ \Phi_{1m} \\ \Phi_3 \\ \Phi_{2p} \\ \Phi_{2m} \end{array} \right) = 
\left(\begin{array}{ccccc} 
1 & 1 & 0 & 0 & 0 \\
-1 & 1 & 0 & 0 & 0 \\
0 & 0 & 1 & 0 & 0 \\
0 & 0 & 0 & 1 & 1 \\
0 & 0 & 0 & 1 & -1
\end{array} \right)
\left(\begin{array}{ccccc} \Phi_1 \\ \Phi_2 \\ \Phi_3 \\ \Phi_4 \\ \Phi_5 \end{array} \right).
\end{split}
\end{equation}
The reason for executing such transformations is that the modes presented by the flux variables $\Phi_{1/2p}$ are free modes with no potential energies (Josephson inductance energies), hence we just need to consider the qubit nodes represented by flux variables $\Phi_{1/2m}$. The capacitance $\mathbf{C}$ in Eq.~(\ref{eq: CapacitanceMatrix}) can be transformed with $\mathbf{\tilde{C}} = \mathbf{(S^{-1})^{T}} \mathbf{C} \mathbf{S^{-1}}$ and the circuit Hamiltonian of the system is given by $H=\frac{1}{2}\mathbf{Q^T} \mathbf{C^{-1}} \mathbf{Q} + U$. For simplicity, we now ignore the free mode variables $\Phi_{1/2p}$ and $\Phi_3$, the reduced capacitance with the inverse matrix form can then be written as
\begin{equation}\label{eq: AMatrix}
\begin{split}
\left(\begin{array}{cc} \dot{\Phi}_{1m} \\ \dot{\Phi}_{2m} \end{array} \right) = 
\left(\begin{array}{cc} 
A_{11} & A_{12} \\
A_{21} & A_{22} 
\end{array} \right)
\left(\begin{array}{cc} Q_{1m} \\ Q_{2m} \end{array} \right),
\end{split}
\end{equation}
where the matrix $\mathbf{A}=1/\mathbf{\tilde{C}}$. Notice that the subscript $1, \, 2$ now represents the subscript $1m, \, 2m$ for qubit modes. Substituting Eq.~(\ref{eq: AMatrix}) into the Hamiltonian, we can acquire the expression as
\begin{equation}\label{eq: Hamiltonian}
\begin{split}
H &= \sum_{i=1,2} \left[\frac{1}{2} A_{ii} Q_{im}^2 - E_{Ji}\cos(\phi_i+\phi_{0i})\right] \\
&+ \sum_{i=1,2; \, i \neq j} \frac{1}{2} A_{ij} Q_{im} Q_{jm}.
\end{split}
\end{equation}
To further characterize the anharmonicities and coupling strength for qubits, we apply canonical quantization procedure to Eq.~(\ref{eq: Hamiltonian}) and obtain
\begin{equation}\label{eq: Hamiltonian2}
\begin{split}
H &= 4E_{C1}\hat{n}_1^2 + 4E_{C2}\hat{n}_2^2 + 4E_{12}\hat{n}_1\hat{n}_2 \\
&- E_{J1}\cos(\hat{\phi}_1 + \phi_{01}) - E_{J2}\cos(\hat{\phi}_2 + \phi_{02}),
\end{split}
\end{equation}
where $\hat{n}_i=-i\partial/\partial\phi_{i}$ ($i=1, \, 2$) are the Copper-pair number operator; $\hat{\phi}_i$ ($i=1, \, 2$) are conjugate operators of $\hat{n}_i$; $E_{Ci}=e^2 A_{ii}/2$ ($i=1, \, 2$) represent the charging energy with qubit anharmonicities $\alpha_i \approx -E_{Ci}$ while $E_{12}=e^2 A_{12}/2$ represents the coupling term in system Hamiltonian.

Next, a second quantization is introduced with defining annihilation and creation operators $\hat{a}_i, \, \hat{a}_i^{\dagger}$ ($i=1, \, 2$) for the qubit modes ($\left[\hat{a}_i, \hat{a}_i^{\dagger} \right]=1$), i.e.,
\begin{equation}\label{eq: AnnihilationOperator}
\begin{split}
\hat{n}_i &= i n_i^{zpf} \left(\hat{a}_i^{\dagger} - \hat{a}_i \right) \\
\hat{\phi}_i &= i \phi_{i}^{zpf} \left(\hat{a}_i^{\dagger} + \hat{a}_i \right),
\end{split}
\end{equation}
where $n_i^{zpf}, \, \phi_{i}^{zpf}$ are the zero-point fluctuations given by $n_i^{zpf}=\frac{1}{\sqrt{2}}\left(\frac{E_{Ji}}{8E_{Ci}} \right)^{\frac{1}{4}}$ and $\phi_i^{zpf}=\frac{1}{\sqrt{2}}\left(\frac{8E_{Ci}}{E_{Ji}} \right)^{\frac{1}{4}}$ ($i=1, \, 2$). Substituting Eq.~(\ref{eq: AnnihilationOperator}) into Eq.~(\ref{eq: Hamiltonian2}) with expanding the cosine terms up to the fourth order, we obtain the Hamiltonian expression
\begin{equation}\label{eq: QubitHamiltonian}
\begin{split}
H &= \left[\omega_{1} + \frac{E_{C1}}{2}\left(1+\frac{\xi_1}{4} \right) - \frac{E_{C1}}{2}\left(1+\frac{9\xi_1}{16} \right) a_1^{\dagger} a_1 \right] a_1^{\dagger} a_1 \\
&+ \left[\omega_{2} + \frac{E_{C2}}{2}\left(1+\frac{\xi_2}{4} \right) - \frac{E_{C2}}{2}\left(1+\frac{9\xi_2}{16} \right) a_2^{\dagger} a_2 \right] a_2^{\dagger} a_2 \\
&+ J_{12}\left(a_1^{\dagger} - a_1 \right)\left(a_2^{\dagger} + a_2 \right),
\end{split}
\end{equation}
where $\omega_i = \sqrt{8E_{Ji}E_{Ci}} - E_{Ci}\left(1+\frac{\xi_i}{4} \right)$ ($i=1, \, 2$) are the frequencies of qubits; $\xi_i = \sqrt{\frac{2E_{Ci}}{E_{Ji}}}$ ($i=1, \, 2$) are the high order coefficient in
the expansion and the effective coupling strength is
\begin{equation}\label{eq: CouplingStrength}
\begin{split}
J_{12} = \frac{E_{12}}{\sqrt{2}}\left(\frac{E_{J1}}{E_{C1}} \frac{E_{J2}}{E_{C2}} \right)^{\frac{1}{4}} \left[1-\frac{1}{8}\left(\xi_1+\xi_2 \right) \right].
\end{split}
\end{equation}

\subsection{Device parameters}

Based on the aforementioned circuit quantization procedure, we have designed the two-qubit transmon qubits coupled through the airbridge, as depicted in Fig. 5(a) in the main text. The corresponding simulation results of the design parameters are documented in Table~\ref{table: DesignParameters}, along with the calculation of qubit characteristic parameters such as anharmonicities and coupling strength.
\begin{table}[ht]
\caption{Design parameters of airbridge-coupled transmon qubits. $C_{\Sigma}$ is the equivalent total capacitance for each qubit; $C_{ab}$ is the capacitance to the ground for the airbridge; $C_{qab}$ is the nearest neighboring coupling capacitance between each qubit and the airbridge capacitance (i.e., $C_{23}$ and $C_{43}$ in Fig.~\ref{fig:FigS1}) with assumption that $C_{13 }=C_{53}=0$ for simplicity; $\omega_i, \, \alpha_i$ ($i=1, \, 2$) represent the frequencies and anharmonicities of qubits and $J_{12}$ represents the effective coupling strength defined in Eq.~(\ref{eq: CouplingStrength}).}
\begin{threeparttable}
\begin{tabular}{cp{2.6cm}<{\centering}p{2.0cm}<{\centering}p{2.0cm}<{\centering}p{2.0cm}<{\centering}}
\\[-7pt]
&\multicolumn{2}{c}{Simulation results} \tabularnewline
\\[-7pt]
\hline
\hline
&{$Q_1$} &{$AB$} &{$Q_2$} \tabularnewline
\hline
$C_{\Sigma}$ (fF) &{$83.8$} &{$\,$}  &{$83.9$}      \tabularnewline
$C_{ab}$ (fF) &{$\,$}  &{$29.5$} &{$\,$}      \tabularnewline
$C_{qab}$ (fF) &{$6.67$} &{$\,$}  &{$6.75$}   \tabularnewline
$\omega_i$ (GHz) &{$4.10$} &{$\,$}  &{$4.10$}  \tabularnewline
$\alpha_i$ (MHz) &{$-240.6$} &{$\,$}  &{$-240.4$}   \tabularnewline
$J_{12}$ (MHz) &{$\,$}  &{$3.32$} &{$\,$}   \tabularnewline
\hline
\end{tabular} \vspace{0pt}

\begin{tabular}{cp{2.6cm}<{\centering}p{2.0cm}<{\centering}p{2.0cm}<{\centering}p{2.0cm}<{\centering}}
\\[-2pt]
&\multicolumn{2}{c}{Experimental results} \tabularnewline
\\[-7pt]
\hline
\hline
&{$Q_1$} &{$AB$} &{$Q_2$} \tabularnewline
\hline
$\omega_i$ (GHz) &{$4.10$}  &{$\,$}  &{$4.54$}  \tabularnewline
$\alpha_i$ (MHz) &{$-240.2$}  &{$\,$}  &{$-234.6$}   \tabularnewline
$J_{12}$ (MHz) &{$\,$}  &{$3.93$}  &{$\,$}   \tabularnewline
\hline
\end{tabular} \vspace{0pt}
\label{table: DesignParameters}
\end{threeparttable}
\end{table}

\section{Airbridge simulation}\label{SecS2}

\begin{figure}
\includegraphics{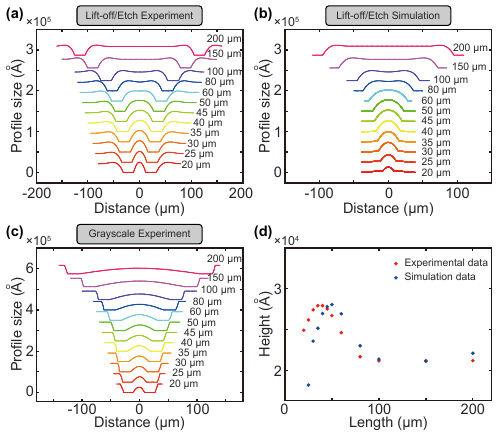}
\caption{(a) Experimental profiles of airbridge scaffolds for etching and lift-off methods. (b) Simulated profiles of scaffold for etching and lift-off methods. (c) Experimental profiles of airbridge scaffold for grayscale lithography method. Notice that curves in (a), (b) and (c) have been shifted for clarity and the length of the airbridges are also marked beside each curves. (d) The height of the airbridges versus the length for both experimental and simulation data.}
\label{fig:FigS2}
\end{figure}

\begin{figure*}
\includegraphics{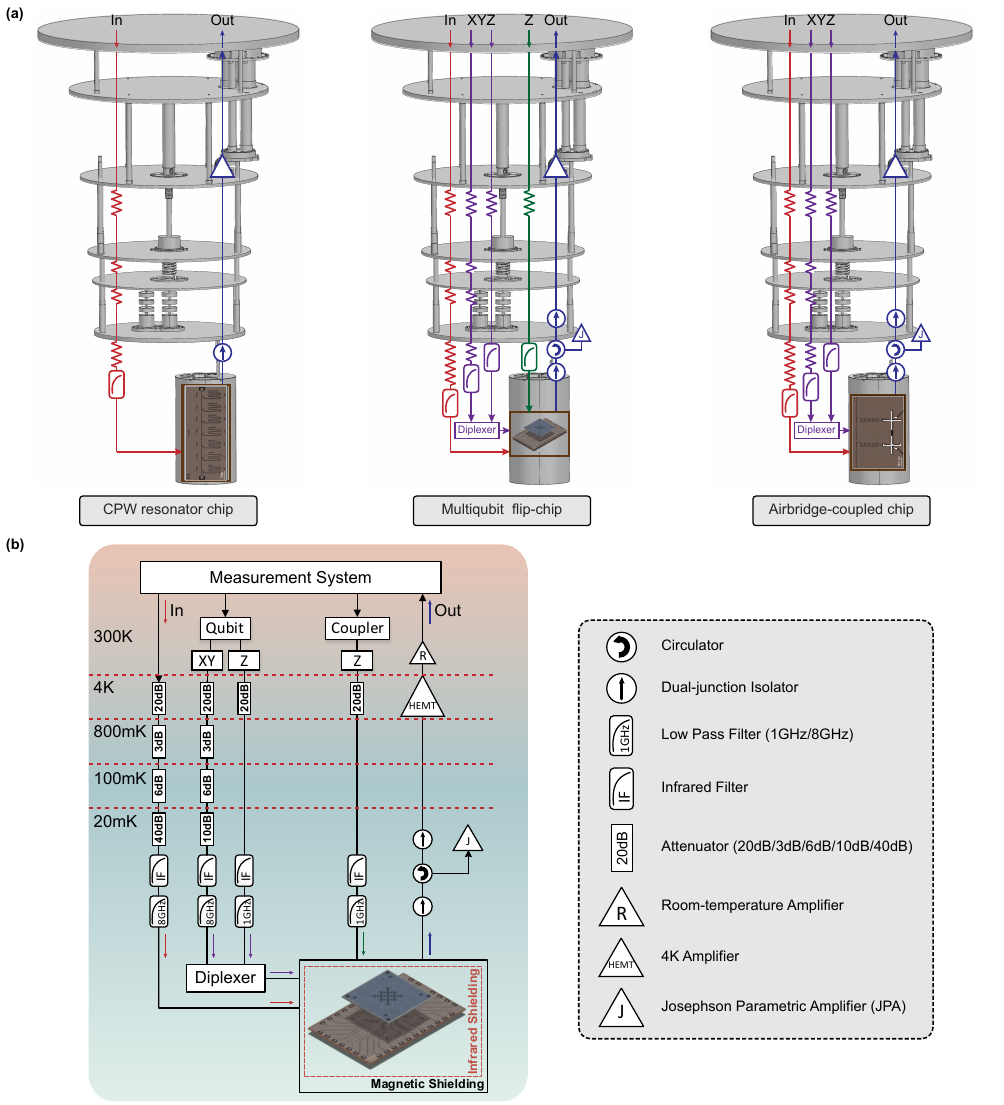}
\caption{Measurement setup. (a) Schematic diagram illustrating the measurement setup employed for various quantum chips utilized in the experiment. Notice that the microwave readout input and output circuits are represented by red and blue lines, while the qubits' $XYZ$ lines and couplers' $Z$ lines are depicted as purple and green lines. (b) The detailed measurement setup for the multiqubit flip-chip in (a) is presented as an exemplification. The gradient color in the background signifies the temperature layer of the dilution refrigerator ranging from room temperature to its lowest point.}
\label{fig:FigS3}
\end{figure*}

We have mentioned in the main text that for the etching and lift-off, the maximum length for a structurally stable airbridge is approximately 60 $\mu$m, while it increases to 200 $\mu$m for airbridges fabricated by grayscale lithography. Hence, we hypothesize that the stability of the airbridge is influenced by the corresponding profile of the photoresist scaffold. To investigate this behavior, we simulate the formation of the scaffold to gain a better understanding. We judge that the arch-shaped scaffold is formed by adding two profiles of the edge scaffold with opposite directions after reflow. Therefore, we extract the edge profile of the photoresist and simulate the profile of the scaffold by adding two edge profiles together. The overlap distance of two profiles represents the length of the airbridge, and the sum height of profiles is the simulated profile of the scaffold. By approaching two edge profiles with different distance, we can then obtain profiles of scaffold with different airbridge lengths.

For the etching and lift-off methods, we observe that an arch-shaped scaffold is formed by combining two edge profiles with opposite directions after undergoing a high-temperature reflow process. By extracting and simulating these edge profiles together, we find that they are nearly the same across different lengths of airbridges. The simulated profiles consistently align with experimental data [Fig.\ref{fig:FigS2}(a)(b)], revealing the appearance of plateaus in scaffolds when exceeding 60 $\mu$m in length. These plateaus could potentially explain why longer airbridges become unstable. Additionally, we extract height information from our simulated scaffolds and qualitatively compare with the experimental data with various lengths, as depicted in Fig.\ref{fig:FigS2}(d). Meanwhile, for grayscale airbridges in Fig.\ref{fig:FigS2}(c), where exposure gray values define their profile precisely, a parabolic shape remains consistent regardless of length.

These results suggest that the edge of the photoresist plays a crucial role in determining scaffold structure and ultimately affects stability of airbridges depending on its characteristics such as having an elongated tail-like feature. During the reflow process, arch-shaped edges are formed. Consequently, by adjusting the reflow temperature, it is possible to systematically modify the profile of edge scaffold. Utilizing simulation described above, one can roughly estimate maximum achievable lengths for stable airbridges after reflow without compromising structural integrity.

\begin{figure*}
\includegraphics{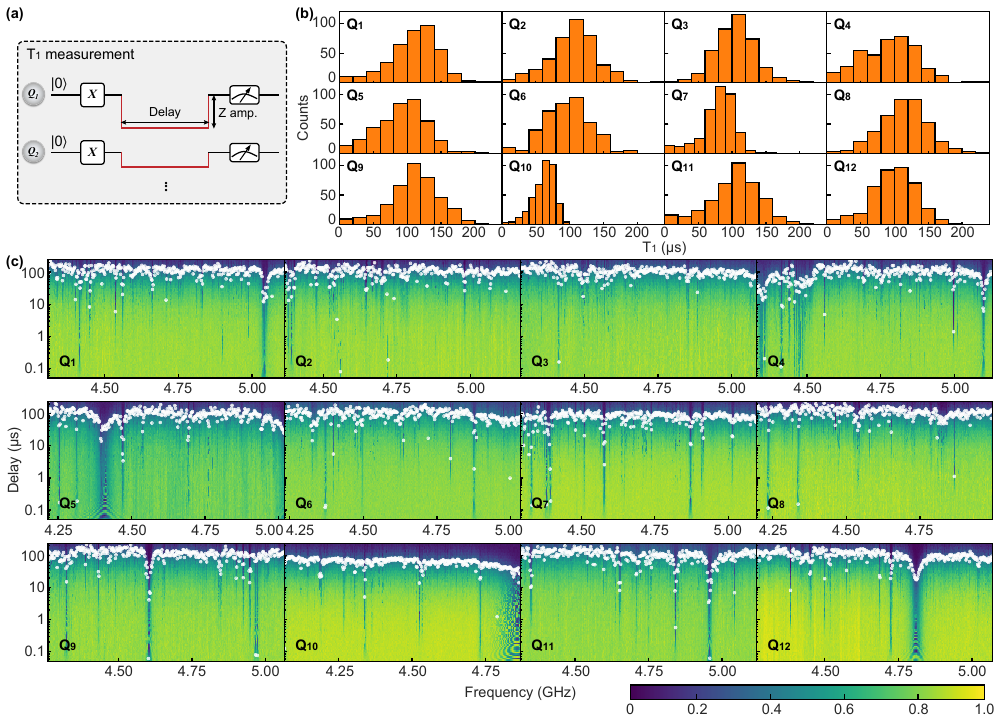}
\caption{(a) A schematic pulse sequence for measuring $T_1$ varied with $Z$ pulse amplitude of qubits. (b) Histogram of the $T_1$ of all qubits extracted from (c) in the multiqubit flip-chip. (c) $T_1$ measurement and the corresponding frequency spectrum (top panel of each images) varied with $Z$ pulse amplitude of qubits. Notice that data of $Q_{13}$ is illustrated in the main text.}
\label{fig:FigS4}
\end{figure*}

\section{Fabrication procedure of the flip-chip processor}\label{SecS3}

The surface-13 flip-chip processor was fabricated using the flip-chip architecture, wherein the qubits and readout resonators/control lines were implemented on separate chips. The top chip consisted of Josephson junctions and shunt capacitances, with the latter being created through etching a 200 nm thick $\alpha$ tantalum film on a sapphire substrate~\cite{rosenberg20173d}. The Josephson junctions were fabricated with the Manhattan style in a double angular evaporation system~\cite{muthusubramanian2023wafer, kreikebaum2020improving}. On the other hand, the bottom chip comprised readout resonators and control lines, which were also formed by etching a 200 nm thick $\alpha$ tantalum film on a sapphire substrate~\cite{place2021new, wang2022towards}. To minimize the crosstalk, airbridges with full-capped structure were incorporated over each control lines, while airbridges with separate structure were utilized for readout lines and resonators. These airbridges were produced using the lift-off method proposed in this study. Additionally, dense indium bumps with 9 $\mu$m thick were deposited onto both top and bottom chips before they were flip bonded together.

\section{Experimental setup and extended data}\label{SecS4}

\subsection{Measurement system}

We have fabricated three types of superconducting quantum chips, including CPW resonators, a tantalum-film based flip-chip quantum processor with tantalum airbridges, as well as a two-qubit quantum chip coupled via the tantalum airbridge. These superconducting quantum devices are all mounted in an aluminum sample holder with varying numbers of ports at a base temperature of $\sim 20$ mK in the dilution refrigerator, as illustrated in Fig.~\ref{fig:FigS3}(a). Considering the purposes of these quantum chips, the setups inside the dilution refrigerator also differ. For example, when measuring the effect of the airbridge on the resonator quality factor using CPW resonator chips, it is essential to ensure high readout efficiency. This means that readout at low power should reach single-photon level sensitivity while preventing power saturation at high power levels~\cite{chen2014fabrication}. Therefore, careful design considerations for readout input and output lines are necessary, including attenuation and room temperature amplification. Regarding qubit devices, each temperature layer should provide sufficient attenuation to suppress potential thermal noise; however excessive attenuation may not be conducive to efficient readout. In addition, to protect the qubits from the flux noise and quasiparticle, we use multi-layer shielding such as magnetic shielding and infrared shielding.

The experimental setup for the surface-13 flip-chip processor is depicted in Fig.~\ref{fig:FigS3}(b) as an example of the fundamental configuration. This versatile setup can be easily adapted to other samples by making necessary modifications. The $XY$ signal generation is achieved through the IQ mixing process, while direct $Z$ modulation is implemented using fast analog signals generated by the self-developed Arbitrary Waveform Generators (AWGs)~\cite{zhang2021exploiting}. Additionally, we employ a Josephson junction parametric amplifier (JPA)~\cite{roy2015broadband}, along with a high-electron mobility transistor amplifier at 4 K and room-temperature amplifiers, to amplify the readout signal, enabling simultaneous single-shot readout for all qubits.

\begin{figure}
\includegraphics{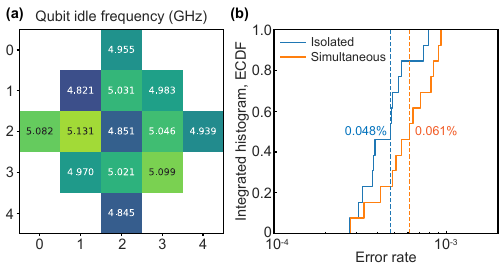}
\caption{(a) Idle frequencies for all qubits when performing parallel single-qubit gate operation. (b) Integrated histogram (empirical cumulative distribution function, ECDF) of single-qubit Pauli errors, measured via isolated (blue) and simultaneous (orange) Randomized Benchmarking. The median of each distribution occurs at 0.5 marked by the dotted line with 0.048\% for isolated-RB and 0.061\% for simultaneous-RB (average values are similar).}
\label{fig:FigS5}
\end{figure}

\subsection{$T_1$ distribution}

The proposed lift-off method offers a robust fabrication technique for the multiqubit quantum processor with long coherence time, utilizing tantalum film as the foundation. In the main text, we present the median $T_1$ distribution for all qubits achieved through frequency tuning with pulse sequence shown in Fig.~\ref{fig:FigS4}(a). Figure.~\ref{fig:FigS4}(b)(c) displays both the $T_1$ distribution of all thirteen qubits varied with frequencies and their corresponding histogram of $T_1$ values. Notably, most qubits exhibit similar median $T_1$ values; however, encountering TLSs leads to a significant drop in $T_1$, which should be circumvented during experimental procedures.

\subsection{Parallel single-qubit gate operations}

Quantum circuits can be executed more efficiently by running quantum gates in parallel. However, this parallel quantum computation is vulnerable to correlated errors such as microwave crosstalk and stray coupling~\cite{google2023suppressing, arute2019quantum}. To assess the effectiveness of tantalum airbridges and tunable couplers in mitigating these errors, we benchmark the single-qubit gate performance using both isolated and simultaneous Randomized Benchmarking (RB)~\cite{arute2019quantum}, as depicted in Fig.~\ref{fig:FigS5}(b). The idle frequencies of all thirteen qubits are also presented in Fig.~\ref{fig:FigS5}(a) for reference. Our result demonstrates a small increase (0.013\%) in single-qubit gate error probabilities, indicating that our device exhibits low microwave crosstalk and stray coupling.


%

\bigskip